\documentclass[12pt]{article}
\usepackage{amssymb}
\usepackage{amsmath}
\usepackage{amsfonts}

\begin{document}

\title{On the problem of quantum control in infinite dimensions}
\author{R. Vilela Mendes\thanks{%
also at IPFN, EURATOM/IST Association; vilela@cii.fc.ul.pt} \\
{\small Centro de Matem\'{a}tica e Aplica\c{c}\~{o}es Fundamentais,}\\
{\small Av. Prof. Gama Pinto, 2, 1649-003 Lisboa, Portugal} \and Vladimir I.
Man'ko\thanks{%
manko@sci.lebedev.ru} \\
{\small P. N. Lebedev Physical Institute }\\
{\small Leninsky Prospect 53, 117924 Moscow, Russia}}
\date{ }
\maketitle

\begin{abstract}
In the framework of bilinear control of the Schr\"{o}dinger equation it has
been proved that the reachable set has a dense complemement in $\mathcal{S}%
\cap \mathcal{H}^{2}$. Hence, in this setting, exact quantum control in
infinite dimensions is not possible. On the other hand it is known that
there is a simple choice of operators which, when applied to an arbitrary
state, generate dense orbits in Hilbert space. Compatibility of these two
results is established in this paper and, in particular, it is proved that
the closure of the reachable set of bilinear control is dense in $\mathcal{S}%
\cap \mathcal{H}^{2}$. The requirements for controllability in infinite
dimensions are also related to the properties of the infinite dimensional
unitary group.
\end{abstract}



\section{Introduction}

The problem of controllability of quantum systems in finite dimensions has
been settled in many papers (see for example \cite{Clark} \cite{Ramakrishna} 
\cite{Fu} \cite{Altafini} \cite{Turinici1} \cite{Albertini}). In contrast,
for infinite-dimensional quantum systems, a few questions are still open 
\cite{Illner} \cite{Tarn}.

In the framework of bilinear control%
\begin{equation}
i\frac{\partial }{\partial t}\psi \left( x,t\right) =\left( H_{0}+g\left(
t\right) B\right) \psi \left( x,t\right)  \label{1.1}
\end{equation}%
with $g\left( t\right) \in L^{2}\left( \left[ 0,T\right] \right) $ and
operators such that $H_{0}$ generates a continuous semigroup and $B$ is bounded,
Turinici \cite{Turinici2} has adapted a result of Ball-Marsden-Slemrod \cite%
{Ball} to show that the set of reachable states from any $\psi _{0}\in 
\mathcal{S}\cap \mathcal{H}^{2}$ has a dense complement in $\mathcal{S}\cap 
\mathcal{H}^{2}$, $\mathcal{S}$ being the Hilbert sphere and $\mathcal{H}%
^{2} $ the $W^{2,2}$ Sobolev space. This is a very general result that
applies whenever the operators in (\ref{1.1}) generate a piecewise (in time)
countable sequence of continuous evolution operators. Then, because
continuous maps map compact sets into compact sets, the reachable set is a
countable union of compact sets. In infinite-dimensional complete metric
spaces, compact sets are nowhere dense hence, by Baire's theorem, the
reachable set is a first category set with dense complement. Therefore exact
bilinear controllability cannot be achieved in $\mathcal{S}\cap \mathcal{H}%
^{2}$.

Several ways may be devised to go beyond this result. Compactness is an
internal property of sets but nowhere density is not, it depends on the
ambient space. Therefore, for example in a higher regularity space, exact
controllability might be achieved. This is the situation in the local
controllability results \cite{Beauchard1} \cite{Beauchard2} in $\mathcal{S}%
\cap \mathcal{H}^{7}$. Another, less explored possibility, would be to
choose a control operator $B$ that does not generate a continuous evolution
operator.

However, what is really important from the physical point of view, is not
exact but approximate controllability, that is, the possibility to approach
any target state with arbitrary accuracy. In the bilinear control setting in 
$\mathcal{S}\cap \mathcal{H}^{2}$ this would correspond to prove that the
reachable set is dense in $\mathcal{S}\cap \mathcal{H}^{2}$. This is likely
to happen because the closure of a first category set is not in general of
first category. In fact, the closure of a linear set is of first category if
and only if it is itself nowhere dense.

Results on approximate controllability in infinite dimensions already exist
in particular cases or imposing some restrictions on the $H_{0}$ and $B$
operators or on their domains (\cite{Nersesyan} \cite{Mirrahimi} \cite%
{Ervedoza} \cite{Mason1} \cite{Mason2}). For example, the exact
controllability in the $\mathcal{H}^{7}$ Sobolev space for a 1D potential,
in \cite{Beauchard1} \cite{Beauchard2}, implies approximate controllability
in $\mathcal{L}^{2}$. In \cite{Mirrahimi} the spectrum is considered to have
only finitely many discrete eigenvalues and in \cite{Nersesyan} the domain
must be bounded. In \cite{Mason1} aproximate controllability requires the
spectrum of $H_{0}$ to be non-resonant and the potential $B$ to couple
directly or indirectly every pair of eigenvectors of $H_{0}$. However these
conditions were later shown \cite{Mason2} to be generic in some sense.

Also, the authors in \cite{Rangan} developed the notion of finitely
controllable infinite dimensional systems. They consider a nested set of
finite-dimensional subspaces of Hilbert space of which the smallest one is
controllable and in each subspace $H_{\alpha }$ acts a set $G_{\alpha }$ of
operators such that any orbit generated by $\exp \left( G_{\alpha }\right) $
contains a vector in a lower dimensional subspace. Then they prove that any
vector in one of the finite-dimensional subspaces may be mapped into any
other vector in another finite-dimensional subspace. This is a powerful
result with practical applications but is not infinite-dimensional
controllability. The subtlety of this difference is related to the fact that 
$G_{\infty }$ (Eq.\ref{3.5}) is a proper subgroup of the
infinite-dimensional unitary group (see Sect.3 for details).

Here we follow a different approach. That approximate controllability is
possible in $\mathcal{S}$ had already been proved in \cite{Vilela1} by the
explicit construction of a small set of unitary operators that, operating in
any $\psi _{0}\in \mathcal{S}$, reach an arbitrarily small neighborhood of
any target state $\psi $. This result has been later generalized to open
quantum systems \cite{Vilela2}. However, this does no settle the question of
approximate bilinear controllability because it is not obvious that the
unitary operators used in \cite{Vilela1} can be generated in the setting of
Eq.(\ref{1.1}).

This is one of purposes of this paper, namely to show that, given any
initial and target states $\left( \psi _{0},\psi \right) $ and an
approximation accuracy $\delta $, it is possible to generate by bilinear
control the required evolution. Use will be made of the results in Ref.\cite%
{Vilela1}, which allows to prove infinite-dimensional controllability with
very mild conditions on the free Hamiltonian $H_{0}$. Approximations of the
shift operator play an important role in this construction. Why the shift
operator or some other \textit{essentially infinite dimensional operator} is
essential for control in infinite dimensions is related to the properties of
the infinite dimensional unitary group. This is explained in detail in
Section 3 and an alternative representation of the shift operator is also
described. The role of essentially infinite dimension operators in the
controllability results may also shed some light on the nature of the
operator conditions used in past attempts to prove approximate
controllability in infinite dimensions.

\section{Approximate bilinear control in infinite dimensions}

The set of operators that in \cite{Vilela1} were shown to implement
approximate controllability in infinite dimensions are the operators of an $%
U\left( 2\right) $ group and the shift operator. By the choice of a
countable basis, any separable Hilbert space is shown to be isomorphic to $%
\ell ^{2}\left( \mathbb{Z}\right) $, the space of double-infinite
square-integrable sequences 
\begin{equation}
a=\left\{ \cdots ,a_{-2},a_{-1},a_{0},a_{1},a_{2},\cdots \right\} \in \ell
^{2}\left( \mathbb{Z}\right)  \label{2.1}
\end{equation}%
$\left\vert a\right\vert =\left( \sum_{-\infty }^{\infty }\left\vert
a_{k}\right\vert ^{2}\right) ^{\frac{1}{2}}<\infty $, with basis 
\begin{equation*}
e_{k}=\left\{ \cdots ,0,0,1_{k},0,0,\cdots \right\}
\end{equation*}%
It was in this setting that the results in \cite{Vilela1} were derived, the
shift operator $U_{+}$ being 
\begin{equation}
U_{+}e_{k}=e_{k+1},\qquad k\in \mathbb{Z}  \label{2.2}
\end{equation}%
with inverse 
\begin{equation}
U_{+}^{-1}e_{k}=e_{k-1},\qquad k\in \mathbb{Z}  \label{2.3}
\end{equation}%
and the $U\left( 2\right) $ group operating in the linear space spanned by $%
e_{0}$ and $e_{1}$ and leaving the complementary space unchanged. It was
then shown that once an initial and target states $\left( \psi _{0},\psi
\right) $ and an accuracy $\delta $ are defined, one may, by the application
of these operators go, in a finite number of steps, from $\psi _{0}$ to a $%
\psi _{n}$ such that $\left\Vert \psi -\psi _{n}\right\Vert <\delta $.

In the space of double infinite sequences, one may choose a representation
Hilbert space $L^{2}(0,2\pi )$, the domain of the operators $H_{0}$ and $B$
being%
\begin{equation*}
D=\left\{ \psi \in \mathcal{H}^{2};\psi \left( 0\right) =\psi \left( 2\pi
\right) \right\}
\end{equation*}%
Then%
\begin{equation*}
\left\{ e_{k}=\frac{1}{\sqrt{2\pi }}e^{ik\theta };k\in \mathbb{Z}\right\}
\end{equation*}%
and the shift operator is%
\begin{equation}
U_{+}=e^{i\theta }  \label{2.4}
\end{equation}

Using this background knowledge we now prove:

\textbf{Proposition:} \textit{Approximate bilinear quantum control, with }$%
H_{0}$\textit{\ generating a strongly continuous semigroup and bounded
control operators }$B$, \textit{is possible in infinite dimensions. That is,
the reachable set is dense in }$S\cap H^{2}$\textit{.}

The proof proceeds by showing that the $U\left( 2\right) $ and the shift
operators may be approximated with arbitrary precision by bounded operators
in the bilinear control context.

Let $U_{1},U_{2},\cdots ,U_{n}$ be the finite set of $U\left( 2\right) $ and
shift operators that take $\psi _{0}$ to a state $\psi _{n}$ through a
sequence of states $\psi _{1}=U_{1}\psi _{0},\psi _{2}=U_{2}\psi _{1},\cdots
\psi _{n}=U_{n}\psi _{n-1}$ such that $\left\Vert \psi -\psi _{n}\right\Vert
<\delta $. Now, considering another set of approximating operators $%
U_{1}^{\prime },U_{2}^{\prime },\cdots ,U_{n}^{\prime }$, and defining $\psi
_{1}^{\prime }=U_{1}^{\prime }\psi _{0},\psi _{2}^{\prime }=U_{2}^{\prime
}\psi _{1}^{\prime },\cdots \psi _{n}^{\prime }=U_{n}^{\prime }\psi
_{n-1}^{\prime }$, we have the following estimate%
\begin{equation*}
\psi _{n}^{\prime }-\psi _{n}=\sum_{k=1}^{n}U_{n}^{\prime }\cdots
U_{n-k+2}^{\prime }\left( U_{n-k+1}^{\prime }-U_{n-k+1}\right) \psi _{n-k}
\end{equation*}%
\begin{equation}
\left\Vert \psi _{n}^{\prime }-\psi _{n}\right\Vert \leq
\sum_{k=1}^{n}\left\Vert \left( U_{n-k+1}^{\prime }-U_{n-k+1}\right) \psi
_{n-k}\right\Vert  \label{2.3a}
\end{equation}

The generator $\theta $ of the shift operator (\ref{2.4}) is not an operator
in $D$ but it can be approximated by bounded operators in $D$. Consider the
following family of bounded operators in $D$%
\begin{equation}
B_{p}\left( \theta \right) =\pi -2\sum_{k=1}^{p}\frac{\sin \left( k\theta
\right) }{k}  \label{2.6}
\end{equation}%
and, for an arbitrary normalized state $\phi =\sum_{k}a_{k}e_{k}$ in $%
L^{2}(0,2\pi )$, compute%
\begin{eqnarray*}
\left\Vert \left( U_{+}-e^{iB_{p}\left( \theta \right) }\right) \phi
\right\Vert ^{2} &=&\frac{1}{2\pi }\left\Vert \sum_{k}a_{k}e^{i\left(
k+1\right) \theta }-\sum_{k}a_{k}e^{i\left( k\theta +B_{p}\left( \theta
\right) \right) }\right\Vert ^{2}= \\
&=&\frac{1}{\pi }\int_{0}^{2\pi }d\theta \sum_{k}\sum_{k^{\prime
}}a_{k}^{\ast }a_{k^{\prime }}e^{-i\left( k-k^{\prime }\right) \theta
}\left( 1-\cos \left( B_{p}\left( \theta \right) -\theta \right) \right) \\
&\leq &\frac{1}{\pi }\left\{ \int_{0}^{2\pi }d\theta \left( 1-\cos \left(
\sum_{k=p+1}^{\infty }\frac{2}{k}\sin \left( k\theta \right) \right) \right)
\right\}
\end{eqnarray*}%
The argument of the cosine in the last term is a Fourier series remainder
which, for $\theta \neq 2\pi $, may be made as small as desired by choosing
a sufficiently large $p$. Because the inequality does not depend on $\phi $
we obtain a norm estimate%
\begin{equation}
\left\Vert U_{+}-e^{iB_{p}\left( \theta \right) }\right\Vert \leq
\varepsilon _{U_{+}}\left( p\right)  \label{2.5}
\end{equation}%
for arbitrarily small $\varepsilon _{U_{+}}\left( p\right) $.

Now, if $H_{0}$ generates a strongly continuous semigroup, $H_{0}+g\left(
t\right) B_{p}\left( \theta \right) $ with $B_{p}\left( \theta \right) $ and 
$g\left( t\right) $ bounded, is also the generator of a strongly continuous
semigroup \cite{perturb}. Then%
\begin{equation*}
\left\Vert \left( e^{iB_{p}\left( \theta \right) }-e^{i\Delta t\left( H_{0}+%
\frac{1}{\Delta t}B_{p}\left( \theta \right) \right) }\right) \psi
\right\Vert \leq \varepsilon _{B}\left( \Delta t,\psi \right)
\end{equation*}%
for any $\psi $, $\varepsilon _{B}\left( \Delta t,\psi \right) $ being as
small as desired by a sufficiently small choice of $\Delta t$.

A similar reasoning applies to a control operator to add to $H_{0}$ to
approximate the $U\left( 2\right) $ transformations to precison $\varepsilon
_{U_{2}}\left( \Delta t,\psi \right) $. Now suppose that to reach $\psi _{n}$
from the initial state $\psi _{0}$ one needs $L$ applications of the shift
operator and $N$ $U\left( 2\right) $ transformations. Then choosing $%
\varepsilon _{U_{+}}\left( p\right) $, $\varepsilon _{B}\left( \Delta t,\psi
\right) $ and $\varepsilon _{U_{2}}\left( \Delta t,\psi \right) $ such that 
\begin{equation*}
L\left( \varepsilon _{U_{+}}\left( p\right) +\varepsilon _{B}\left( \Delta
t,\psi \right) \right) +N\varepsilon _{U_{2}}\left( \Delta t,\psi \right)
\leq \delta
\end{equation*}%
one concludes from (\ref{2.3a}) that the desired control precision is
obtained. This completes the proof.

\section{The shift operator and the infinite dimensional unitary group}

In the proof of approximate controllability in infinite dimensions in \cite%
{Vilela1}, the shift operator played an important role. Of course, the
choice of operators implementing quantum control in infinite dimensions is
not unique, but the fact that an operator with properties similar to the
shift is needed reflects the special features of the infinite-dimensional
unitary group. The infinite dimensional unitary and orthogonal groups, $%
U\left( \infty \right) $ and $O\left( \infty \right) $, are clearly
transitive in complex and real infinite-dimensional Hilbert space. Therefore
the operators that implement control in infinite dimensions must somehow be
able to generate these groups. The suitable mathematical setting for the
groups $U\left( \infty \right) $ or $O\left( \infty \right) $ is a Gelfand
triplet 
\begin{equation}
E^{\ast }\supset L^{2}\left( \mathbb{R}^{d}\right) \supset E  \label{3.1}
\end{equation}%
$E$ being a nuclear space obtained as the limit of a sequence of Hilbert
spaces with successively larger norms. An element $g$ of $U\left( \infty
\right) $ is a transformation in $E$ such that 
\begin{equation}
\left\Vert g\xi \right\Vert =\left\Vert \xi \right\Vert  \label{3.2}
\end{equation}%
By duality $\left\langle x,g\xi \right\rangle =\left\langle g^{\ast }x,\xi
\right\rangle $, $x\in E^{\ast },\xi \in E$, the infinite-dimensional
unitary group is also defined on $E^{\ast }$, the two groups being
algebraically isomorphic.

For the harmonic analysis on $U\left( \infty \right) $ one needs functionals
on $E^{\ast }$. $U\left( \infty \right) $ is a complexification of $O\left(
\infty \right) $ and a standard result states that if a measure $\mu $ is
invariant under $O\left( \infty \right) $ it must be of the form 
\begin{equation*}
\mu =a\delta _{0}+\int \mu _{\sigma }dm\left( \sigma \right)
\end{equation*}%
a sum of a delta and Gaussian measures $\mu _{\sigma }$ with variance $%
\sigma ^{2}$. Hence we are led to consider the $\left( L^{2}\right) $ space
of functionals on $E^{\ast }$ with a $O\left( \infty \right) -$invariant
Gaussian measure 
\begin{equation*}
\left( L^{2}\right) =L^{2}(E^{\ast },B,\mu )
\end{equation*}%
$B$ being generated by the cylinder sets in $E^{\ast }$ and $\mu $ the
measure with characteristic functional 
\begin{equation*}
C\left( f\right) =\int_{S^{\ast }}e^{i\left\langle x,f\right\rangle }d\mu
\left( x\right) =e^{-\frac{1}{2}\left\Vert f\right\Vert ^{2}},\hspace{1cm}%
x\in E^{\ast },f\in E
\end{equation*}

In conclusion: the proper framework to study transitive actions and
functional analysis in infinite dimensional quantum spaces is the complex
white noise setting \cite{Hida1}. In this context many useful results are
already available. For example, the regular representation of $U\left(
\infty \right) $%
\begin{equation*}
U_{g}\varphi \left( z\right) =\varphi \left( g^{\ast }z\right) ,\hspace{1cm}%
z\in E_{c}^{\ast },\varphi \in \left( L_{c}^{2}\right) \cong \left(
L^{2}\right) \otimes \left( L^{2}\right)
\end{equation*}%
splits into irreducible representations \cite{Okamoto} corresponding to the
Fock space (chaos expansion) decomposition of $\left( L_{c}^{2}\right) $%
\begin{equation*}
\left( L^{2}\right) =\oplus _{n=0}^{\infty }\left( \oplus
_{k=0}^{n}H_{n-k,k}\right)
\end{equation*}%
$H_{n-k,k}$ being a complex Fourier-Hermite polynomial of degree $\left(
n-k\right) $ in $\left\langle z,\xi \right\rangle $ and of degree $k$ in $%
\left\langle \overset{\_}{z},\overset{\_}{\xi }\right\rangle $

Furthermore, results concerning a classification of the subgroups of $%
U\left( \infty \right) $ are useful to understand the requirements of
quantum control in infinite dimensions. In particular one must distinguish
between subgroups that only involve transformations that may be approximated
by finite-dimensional transformations like $G_{\infty }$, obtained as the
limit of a sequence of finite-dimensional unitary groups 
\begin{equation}
G_{n}=\left\{ g\in U\left( \infty \right) ,\left. g\right\vert _{V_{n}}\in
U\left( n\right) ,\left. g\right\vert _{V_{n}^{\bot }}=I\right\}  \label{3.4}
\end{equation}%
\begin{equation}
G_{\infty }=\text{proj}\lim_{n\rightarrow \infty }G_{n}  \label{3.5}
\end{equation}%
from those that contain transformations changing, in a significant way,
infinitely many coordinates. These group elements are called \textit{%
essentially infinite-dimensional} (see the section 4 for a definition). The
essential point to remember is that to generate $U\left( \infty \right) $,
and therefore to be transitive in infinite dimensions, some essentially
infinite dimensional elements are needed. This is why the shift operator or
some other essentially infinite-dimensional operation is required for
control in $\mathcal{S}\cap \mathcal{H}^{2}$.

In our mathematical construction we have represented a separable Hilbert as
a space of double-infinite sequences. Given the importance of essentially
infinite-dimensional operators for the quantum control in $\mathcal{S}\cap 
\mathcal{H}^{2}$ we include in the next section an implementation of the
shift operator in an oscillator-like basis, which may be closer to the usual
physical applications.

\section{The shift operator in an oscillator-like basis}

In the Gelfand triplet setting%
\begin{equation*}
E\subset H\subset E^{\ast }
\end{equation*}%
with the white noise measure $\mu $ in $E^{\ast }$, choose an orthonormal
basis in $E$%
\begin{equation*}
\left\{ \xi _{i}:i=0,1,2,\cdots \right\}
\end{equation*}%
In this basis one has the usual raising and lowering operators $a^{+}$ and $%
a $ and define the operators%
\begin{eqnarray*}
A^{+} &:&A^{+}\xi _{i}=\xi _{i+1} \\
A &:&%
\begin{array}{l}
A\xi _{i}=\xi _{i-1};i\neq 0 \\ 
A\xi _{0}=0%
\end{array}%
\end{eqnarray*}%
that is, $A^{+}=$ $a^{+}\frac{1}{\sqrt{a^{+}a+1}}$ and $A=\frac{1}{\sqrt{%
a^{+}a+1}}a$.

The projection operator $P_{0}$ on the basis state $\xi _{0}$ is%
\begin{equation*}
P_{0}=\left\vert \xi _{0}\right\rangle \left\langle \xi _{0}\right\vert
=1-A^{+}A
\end{equation*}%
and in any other state $\xi _{n}$ is%
\begin{equation*}
P_{n}=\left\vert \xi _{n}\right\rangle \left\langle \xi _{n}\right\vert
=\left( A^{+}\right) ^{n}P_{0}\left( A\right) ^{n}
\end{equation*}%
The elementary operator $P_{jk}$ that transforms $\xi _{k}$ into $\xi _{j}$
is%
\begin{equation*}
P_{jk}=\left\vert \xi _{j}\right\rangle \left\langle \xi _{k}\right\vert
=\left( A^{+}\right) ^{j}P_{0}\left( A\right) ^{k}
\end{equation*}%
and one also define the following parity operators%
\begin{equation*}
P_{\pm }=\frac{1}{2}\left( 1\pm e^{i\pi a^{+}a}\right)
\end{equation*}%
Now the operator%
\begin{equation*}
U_{+}=\left( A^{+}\right) ^{2}P_{+}+\left( A\right) ^{2}P_{-}+P_{01}
\end{equation*}%
plays the same role as the shift operator in the space of double-infinite
square-integrable sequences, as may easily be seen by the appropriate
renumbering of a double infinite sequence. Notice that this operator is
different from the resonant driving field $\left( u\left( t\right) x\right) $%
, used in the discussions of controllability of the harmonic oscillator \cite%
{Kime} \ \cite{Rouchon}, which together with free Hamiltonian generates a
four-dimensional Lie algebra. $U_{+}$ together with an $SU\left( 2\right) $
group acting in the subspace $\left\{ \xi _{0},\xi _{2}\right\} $ generates
an infinite-dimensional Lie algebra and controllability in infinite
dimensions is obtained. $U_{+}$ is an essentially infinite dimensional
operator. This notion is rigorously defined through the \textit{average
power }\cite{Hida1}, that is,%
\begin{equation*}
ap\left( U_{+}\right) \left( x\right) =\lim \sup_{N\rightarrow \infty }\frac{%
1}{N}\sum_{n=1}^{N}\left\langle x,U_{+}\xi _{n}-\xi _{n}\right\rangle ^{2}
\end{equation*}%
$x\in E^{\ast }$. If $ap\left( U\right) \left( x\right) $ for an operator $U$
is positive almost surely for the measure $\mu $ in $E^{\ast }$, the
operator is called \textit{essentially infinite-dimensional}. Qualitatively
it means that it acts, in a significant way, in infinitely many coordinates.
In the opposite case, if $ap\left( U\right) \left( x\right) =0$ almost
surely, then $U$ may be approximated by transformations acting on finite
dimensional subspaces. The average power of $U_{+}$ is $2$ almost surely.

Other essentially infinite-dimensional operators may be obtained by
constructions similar to the one used for $U_{+}$. As follows from the
nature of the infinite-dimensional unitary group, at least one such operator
(or an arbitrarily close approximation there of) is needed to obtain density
in $\mathcal{S}\cap \mathcal{H}^{2}$.

\section{Conclusions}

In addition to establishingh that under mild conditions the closure of the
reachable set of bilinear control is dense in $\mathcal{S}\cap \mathcal{H}%
^{2}$, we have also put in evidence the special role of essentially
infinite-dimensional operators in quantum control. 

The central role here was played by the shift operator and approximations
thereof. This is an operator that behaves like the application of a magnetic
field pulse to a charged particle in a circle (a charged plane rotator),
which shifts the eigenstates one level up. Other simple essentially
infinite-dimensional operators are described in Ref.\cite{Hida1}, which may
be used as a guide to develop control methodologies for infinite-dimensional
quantum systems.

\end{document}